\documentclass[preprintnumbers,article,amsmath,amssymb,floatfix,10pt,prd,onecolumn,
superscriptaddress,nofootinbib]{revtex4}
\usepackage[colorlinks=true, pdfstartview=FitV, linkcolor=blue, citecolor=red, urlcolor=magenta]{hyperref}
\usepackage{bbm}
\usepackage{amsfonts}
\usepackage{mathrsfs}
\usepackage{latexsym}
\usepackage{epsfig}
\usepackage{epstopdf}
\usepackage{epstopdf}
\usepackage{graphicx}
\usepackage{amssymb}
\usepackage{amsmath}
\usepackage{dcolumn}
\usepackage{bm}
\usepackage{color}
\usepackage{comment}
\usepackage{xcolor}
\begin{document}
\title{Thermal Fluctuations Evolution of the New Schwarzschild Black Hole}

\author{Zunaira Akhtar}
\email{zunaira.math.pu@gmail.com}
\affiliation{Department of Mathematics, University of the Punjab, Quaid-e-Azam Campus, Lahore 54590, Pakistan}

\author{Rimsha Babar}
\email{rimsha.babar10@gmail.com}
\affiliation{Division of Science and Technology, University of Education, Township, Lahore 54590, Pakistan}

\author{Riasat Ali}
\email{riasatyasin@gmail.com}
\affiliation{Department of Mathematics, GC
University Faisalabad Layyah Campus, Layyah-31200, Pakistan}

\begin{abstract}
We study the thermodynamic analysis and logarithm corrections of the new Schwarzschild black hole. We compute the thermodynamic quantities like entropy, Hawking temperature and heat capacity.
The area of black holes never decreases because they absorb everything from their surroundings due to high gravity.
In this regard, the area-entropy relation proposed by Bekenstein needs to be corrected, leading to the concept
of logarithmic corrections. To do so, we obtain the corrected entropy for new Schwarzschild black hole to analyze the effects of thermal fluctuations and we evaluate the thermodynamic quantities like specific heat, internal energy, Helmholtz free energy, Gibbs free energy, enthalpy and pressure in the presence of correction parameter $\eta$. Furthermore, we check the stability of the system with the help of heat capacity and well known Hessian matrix technique. By our graphical analysis, we observe that the thermal fluctuations effects the stability of small radii black holes (e.g., New Schwarzchild black hole) and therefore, small black holes get unstable regions due to these first-order corrections.
\end{abstract}
\maketitle

\date{\today}


\section{Introduction}
A theory based on black hole (BH) thermodynamics was developed in $1970$ by taking into account the fundamental laws of thermodynamics to investigate the BHs in the field of general relativity (GR).
In spite of the fact that this theory has not yet been clearly understood and figured out, the presence of BH thermodynamics firmly proposes the profound and crucial connection between GR, thermodynamics and quantum field theory (QFT). 
Although, it seems to be based on the fundamental principles of thermodynamics and explain the behavior of BHs under the limitations of the laws of thermodynamics with classical and semi-classical reasoning, its significance goes far beyond the relationship between thermodynamics and BHs. The fundamental relationship between BHs and thermodynamics were at first introduced by Hawking and Bekenstein \cite{[1]}.
Black holes act as thermodynamic objects that transmit radiation through horizon by utilizing the QFT in curved space–time, called Hawking radiation with the properties, temperature corresponding to surface gravity at the horizon and entropy associates to one quarter region of the horizon. Although, the fundamental laws of thermodynamics have close connection with BHs physics e.g., the $2$nd law of thermodynamic is similar to the $2$nd law of BH dynamics (the law of area), that means the surface of a BH never decrease and the mass of the BH, Hawking temperature and entropy fulfill the requirements of the first law of thermodynamics \cite{[1a]}. Schwarzschild ($1916$) \cite{[6]} proposed a spherically symmetric solution for matter, the "internal Schwarzschild solution", which depends upon the Einstein's field equations. In $1916$, Reissner \cite{[7]}, presented a modified form of the Schwarzschild BH by adding charge parameter and this work was subsequently finalized by Weyl \cite{[8]} and Nordstr\"{o}m \cite{[9]} and therefore, known as Reissner-Nordstr\"{o}m solution.
Hawking ($1974$) reported a theoretical derivation of BH radiation \cite{[2]}. 
A Schwarzschild BH of mass $M$ transmits thermal radiation at the temperature
\begin{equation} \label{dd}
T_{h} = \frac{\hbar}{8\pi kGM},
\end{equation}
here $k$ is the Boltzmann
constant. Moreover, the  Bekenstein entropy of a Schwarzschild BH Under the Hawking’s outcome is given as \cite{[1b]}
\begin{equation}
S =\frac{k c^3 A}{4\hbar G},
\end{equation}
here $A$ represents the area of BH. Hawking’s important result isn't explicitly identical to quantum gravity, it may be a capable application of quantum field theory in curved space-time but incorporates an exceptionally strong affect on the field of quantum gravity. It opens another field of exploration (i.e., BH thermodynamics) and it highlights the quantum gravitational issues of understanding the statistical basic of the entropy. Black hole thermodynamics is one of the most fascinating thing in present age and an important area of study that looks to accommodate the laws of thermodynamics through BH event horizon that is broadly examined in the literature. The metric and thermodynamics of Schwarschild BH has been widely discussed in different aspects. Andre and Lemos \cite{[3]} have investigated the thermodynamics of $d$-dimensional Schwarzschild BH i.e., Hawking temperature, Bekenstein entropy and heat capacity as well as they have also discussed their thermal stability. The stability and non-physical
thermodynamic conditions of a $d$-dimensional Schwarzschild
BH have been analyzed very deeply \cite{[10]}-\cite{[12]}.
Nicolini and his colleagues \cite{[13]} have studied the noncommutative radiating Schwarzschild BH behavior and discussed the evaporation conditions with the analysis of Hawking temperature via horizon.
Miao and Wu \cite{[14]} have investigated Schwarzschild-AdS BH thermodynamics with a negligible length and calculated the Hawking temperature and entropy of BH and discussed the phase transitions with various type of vacuum pressure from self-regular
Schwarzschild-AdS BH.
The thermodynamics of BH surrounded by quintessence or without quintessence under generalized uncertainty principle effects have been discussed \cite{R12, [4], R11}. 

Ghaderi and Malakolkalami \cite{[5]} have also studied the thermodynamics of the Schwarzschild BH and derived the temperature, heat capacity and mass associated with entropy. Chen and Zheng \cite{1} analyized the thermodynamics of new Schwarzschild BH. Additionally, the thermodynamics of variety of BHs is studied by many researchers under the consideration of quantum gravity effects  and consistency of results was made sure by reducing the Schwarzschild Hawking temperature\cite{RA}-\cite{T5}. Sharif and Zunaira \cite{T7,T8} adopted the Weyl corrections concept to investigate the thermal fluctuations of charged BHs, the authors also explored the quasi-normal mode, and it was disclosed that under the consideration of first order correction the system shows unstable behavior for small radii of BHs. However, the system local and global stability was anticipated by heat capacity and Hessian matrix. Furthermore, the phase transition and thermodynamics of Newman–Unti–Tamburino BH were evaluated by consideration of accelerating and rotating charged pairs \cite{T9,T10}. The main purpose of this work is to discuss the 
give the thermodynamic analysis and logarithm corrections of new Schwarzschild BH with the help of graphical interpretation as well to evaluate the corrected entropy to check the effects of thermal fluctuations on the given space-time.
The paper is organized in this way; In Sec. \textbf{II},
we discuss the thermodynamic of new version of Schwarzschild BH. 
Sec. \textbf{III} contains the detailed analysis of Hawking temperature of new Schwarzschild BH.
Sec. \textbf{IV} investigates the corrected entropy to discuss the effects of thermal fluctuations on the considered geometry. Section \textbf{V}, comprises the main results.

\section{Thermodynamic of new Schwarzschild black hole}

This section describes the thermodynamic properties like Bekenstein-entropy, temperature and heat capacity of new version of Schwarzschild BH. Further, we check the stability of the given system by graphical analysis of heat capacity.
The metric of new type of BH is defined as  \cite{1}
\begin{equation}\label{1.1}
ds^{2}=-G(r)dt^{2}+\frac{1}{G(r)}dr^{2}+r^{2}d\Omega^{2},
\end{equation}
here $d\Omega^{2}=d\theta^{2}+\sin^{2}\theta d\phi^{2}$ and
\begin{equation}\label{1.2}
G(r)=1-\frac{\mu}{r}+\frac{p^{2}}{r^{2}},
\end{equation}
where $p$ and $M$ are algebraic thermodynamic parameters and ADM mass of the given geometry, respectively. Here $M=\mu/2$, and BH has two surfaces like radius i.e., inner ($r_{-}$) and outer $r_{+}$ horizon at $G(r)=0$ in this form
\begin{equation}
r_{+}=\frac{1}{2}\Big(\mu+\sqrt{\mu^{2}-4p^2}\Big), ~~~r_{-}=\frac{1}{2}\Big(\mu-\sqrt{\mu^{2}-4p^2}\Big).
\end{equation}
At this condition $r_{+}$=$r_{-}$, the BH becomes extreme BH. Moreover, the algebraic parameter and ADM mass can be expanded like that
\begin{equation}
\mu=r_{-}+r_{+},\quad p^{2}=r_{-}r_{+}.
\end{equation}
Excluding the naked singularity at $r=0$, the condition $p^{2}\leq \mu^{2}/4$ is satisfied. The BH entropy in the form of area is defined as
\begin{equation}\label{1.3}
\bar{S}=\frac{\tilde{A}}{4}=\pi r_{+}^{2}.
\end{equation}
The corresponding temperature of BH horizon is calculated as
\begin{equation}\label{1.4}
T=\Big(\frac{\partial M}{\partial
\tilde{S}}\Big)_{q}=\frac{r_{+}-r_{-}}{4\pi r_{+}^{2}}.
\end{equation}
Moreover, the heat capacity of the system can be calculated by the following expression
\begin{equation}\label{1.5}
C=T\frac{\partial S}{\partial T}=\frac{4 \pi  r_+^2 \left(r_+-r_-\right)}{-2 r_--r_+}.
\end{equation}

By graphical behaviour of metric function in \textbf{Fig. 1} in \textbf{Appendix A}, it is noted that the appearance of naked singularities for the different choices of inner horizon and the geometry of this BH is just looks like Reissner BH \cite{2}-\cite{4}.
The \textbf{Fig. 2} \textbf{Appendix A} shows the Physical behavior of hawking temperature along to horizon radius for different choices of inner horizon. It can be observed that for the larger choices of inner horizon, the graph gradually decreases, means the energy release from the system at this moment.
\begin{center}
\includegraphics[width=7.1cm]{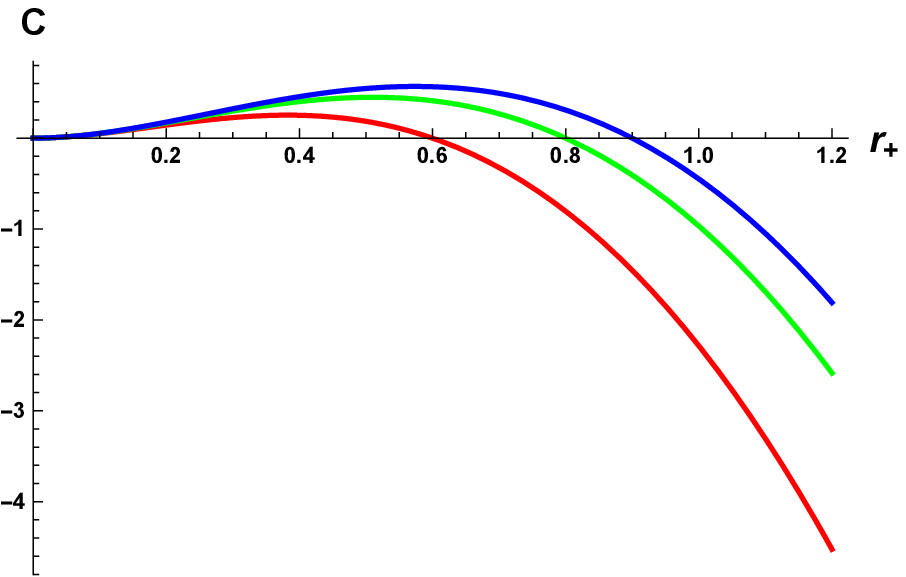}\\
{\textbf{Figure 3}: Heat capacity versus $r_+$. }
\end{center}
Generally, the positive value of heat capacity means the system shows stable behavior while the negative value tends toward unstable one \cite{5}. In \textbf{Fig. $3$}, the graphs shows positive behavior for the small value of horizon radius (stable region) after some time shift towards negative side (unstable region). It can be observed that there exist a phase transition at these points ($0.6, 0.8, 0.9$), that means graph crosses the boundary of horizon radius at these points.

\section{Detailed Analysis of Hawking Temperature for New Schwarzscild BH}

The detailed analysis of Hawking temperature $T_{h}$ for new Schwarzschild BH in terms of algebraic thermodynamic parameter $p$ and $\mu$ is given in this section.
In order to discuss the thermodynamics, the $T_{h}$ can be calculated by the given formula
\begin{equation}\label{cc}
T_{h}=\frac{k_{h}}{2\pi}=\frac{1}{4\pi\sqrt{-g_{00}g_{11}}}\Big|\frac{dg_{00}}{dr}\Big|_{r=r_+},
\end{equation}
here $k_{h}$ represents the surface gravity. By using Eq. (\ref{1.1}) into Eq. (\ref{cc}) , we get
\begin{equation}\label{bb}
T_{h}=\frac{1}{4\pi}\Big[\frac{\mu r_{+}-2p^2}{r_{+}^3}\Big].
\end{equation}
The above expression represents the Hawking temperature of new Schwarzschild BH that depends upon the parameter $\mu$, thermodynamic parameter $p$ and horizon radius $r_{+}$. It is note worthy that by putting $p=0$ and $\mu=2M$ in above Eq. (\ref{bb}), we get the Hawking temperature of Schwarzschild BH given in Eq. (\ref{dd}) for $\hbar=k=G=1$ \cite{[2]}.

\begin{center}
\includegraphics[width=7.1cm]{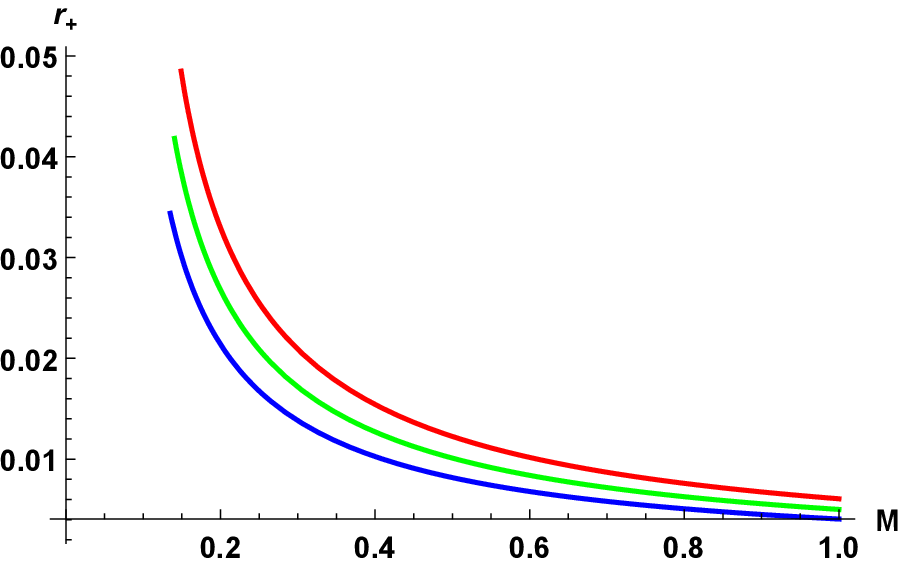}\\
{\textbf{Figure 4}: Horizon $r_+$ versus $M$ for~$p= 0.09$ (blue), $p=0.1$
(green) and $p=0.11$ (red).}
\end{center}

The \textbf{Fig. 4} represents the horizon-mass relation for new Schwarzschild BH. A stable behavior of BH can be observed for various values of $p$ in the region $0\leqslant M\leqslant1$.
The \textbf{Fig. 5} in \textbf{Appendix B} indicates the behavior of temperature for varying values of parameter $\mu$ and fixed value of thermodynamic parameter $p=0.7$. The curves are generic and defined actually by the asymptotic behavior when $T_{h}\rightarrow0$ as $r_+\rightarrow\infty$. At initial the negative temperature shows the non-physical form of BH but after some time BH comes in its stable form and the maximum height of temperature for $r_{+}>0$ (non-zero horizon) indicates the condition at which BH remnant left.
The \textbf{Fig. 6} \textbf{Appendix B} depicts the conduct of temperature versus $r_+$ for various values of thermodynamic parameter $p$ and fixed value of $\mu=50$ in the range $0\leqslant r_+\leqslant15$. At non-zero horizon the temperature shows an extreme points and then eventually drops from this height and goes to attain an asymptotically flat form till $r_+\rightarrow\infty$ that indicates the complete stable form of BH. The decreasing value of $T_{h}$ with increasing value of $r_+$ completely satisfy the Hawking's phenomenon.
\section{Thermal Fluctuations}
This section evaluates the thermal fluctuations of new type of BH. Here, we will compare the behavior of original and corrected thermodynamic quantities like internal energy, entropy, enthalpy, Helmholtz free energy, pressure, specific heat and Gibbs free energy. To check the corrected entropy, the partition function can be given in the form \cite{6}
\begin{equation}\label{2.1}
\tilde{Z}(\beta)=\int_{0}^{\infty} dE \rho(E)e^{-\beta E},
\end{equation}
where the density of state is calculated through inverse Laplace transform. The expression of density of state is given as
\begin{equation}\label{2.2}
\rho(E)=\frac{1}{2\pi
i}\int_{\beta_{0}-i\infty}^{\beta_{0}+i\infty}d\beta \tilde{Z}(\beta)
\exp(\beta E) =\frac{1}{2\pi
i}\int_{\beta_{0}-i\infty}^{\beta_{0}+i\infty}d\beta
\exp(\tilde{S}(\beta)),
\end{equation}
where $\tilde{S}(\beta)=\ln Z(\beta)+\beta E$ is called exact
entropy of system that depends upon Hawking temperature.
With the help of steepest descent method, the entropy is defined by the given formula
\begin{equation}\label{2.3}
\tilde{S}(\beta)=\bar{S}+\frac{1}{2}(\beta-\beta_{0})^{2}
\frac{\partial^{2}\tilde{S}(\beta)}{\partial
\beta^{2}}\Big|_{\beta=\beta_{0}}+\text{higher-order terms},
\end{equation}
under the conditions  $\frac{\partial
\tilde{S}}{\partial\beta}=0$ and $\frac{\partial^{2}
\tilde{S}}{\partial\beta^{2}}>0$, the entropy relation along fluctuations gets the form \cite{7}- \cite{9}
\begin{equation}\label{2.4}
\tilde{S}=\bar{S}-\eta \ln(ST^{2})+\frac{\gamma}{S},
\end{equation}
where $\eta$ and $\gamma$  are correction parameters. There are $4$ different cases for the choices of value of these correction parameter, defined as follows
\begin{itemize}
\item with $ \eta ,~ \gamma \rightarrow 0$, the expression of original entropy is obtained.
\item When $\eta \rightarrow 1,~ \gamma \rightarrow 0$, we can get the first order/logarithmic corrections.
\item For $\eta \rightarrow 0,~\gamma \rightarrow 1$, we can obtained the second order correction terms.
\item For $\eta,~\gamma \rightarrow 1$, higher order
corrections are obtained.
\end{itemize}

By substituting the condition ($\eta\rightarrow 1,~ \gamma \rightarrow 0$) as well as Eq. (\ref{1.3}) and Eq. (\ref{1.4}) into Eq. (\ref{2.4}), we get
\begin{equation}\label{2.5}
\tilde{S}=\pi  r_+^2-\frac{1}{2} \eta  \log \left(\frac{\left(r_--r_+\right){}^2}{16 \pi  r_+^2}\right).
\end{equation}
The second term describes the effects of fluctuations to entropy. 
\begin{center}
\includegraphics[width=7.1cm]{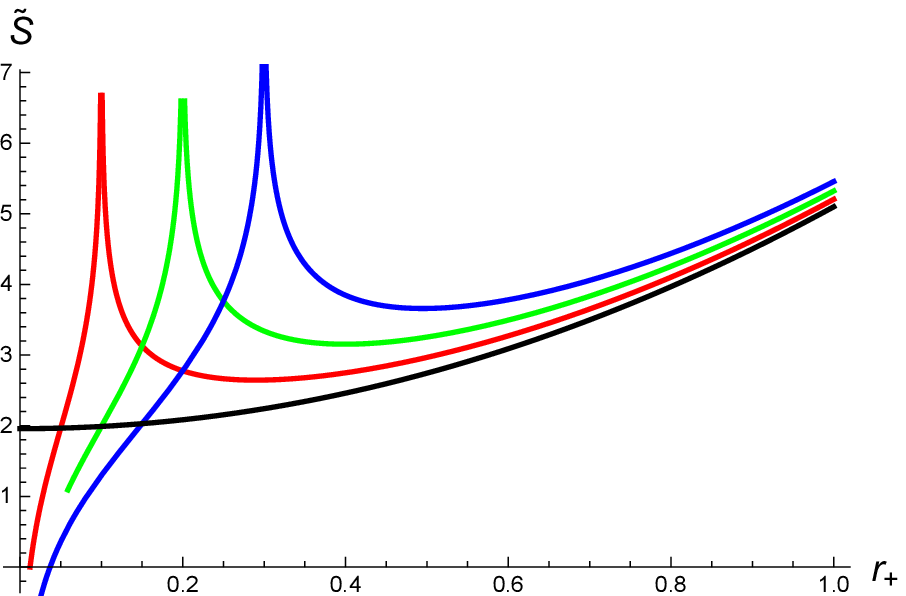}\\
{\textbf{Figure 7}: Entropy versus $r_+$ }
\end{center}
We plot our graphs by taking the different choices of correction parameter $\eta$ = $0.4$ (red), $0.5$ (green), $0.6$ (blue) and $0$ (black). 
The \textbf{Fig. 7} represents that the entropy of the space-time is monotonically increasing throughout the considered domain. The peak of entropy decreases for a specific value of the radii, that means this abrupt change represents a thermal fluctuation in the system. However, it is observed that these fluctuations are more efficient for smaller BHs. Moreover, the $1$st law of BH thermodynamics along thermal fluctuations can be defined as \cite{1}
\begin{eqnarray}\label{2.51}
dM=\mathcal{T}d\tilde{S}+\Phi dp,
\end{eqnarray}
here $\mathcal{T}$ and $\Phi$ represents the corrected temperature as well as electrostatic potential of considered geometry. In order to check the effects of thermal fluctuations upon modified thermodynamic quantities the potential functions can be calculated as
\begin{equation}\nonumber
\mathcal{T}=\Big(\frac{\partial M}{\partial
\tilde{S}}\Big)_{q}=\frac{\eta }{2 r_--2 r_+}+\frac{\eta }{2 r_+}+\pi  r_+ ,\quad \varphi=\Big(\frac{\partial M}{\partial
p}\Big)_{\tilde{S}}=\frac{\sqrt{r_+r_-}}{r_+}=\frac{p}{r_+}.
\end{equation}
By taking into account the corrected entropy and temperature expression, we explore the other thermodynamic quantities under the effects of thermal fluctuations. Thus, the Helmholtz energy ($F=-\int \tilde{S}dT$) leads to the form
\begin{equation}\label{2.6}
F=\Big[-\frac{\eta  r_-}{r_+^2}-\frac{\eta  \left(r_+-r_-\right) \log \left(\frac{\left(r_--r_+\right)^2}{16 \pi  r_+^2}\right)}{r_+^2}-2 \pi  r_++4 \pi r_{-} -\log
   \left(r_+\right)\Big]\Big[8 \pi\Big]^{-1}.
\end{equation}
\begin{center}
\includegraphics[width=7.1cm]{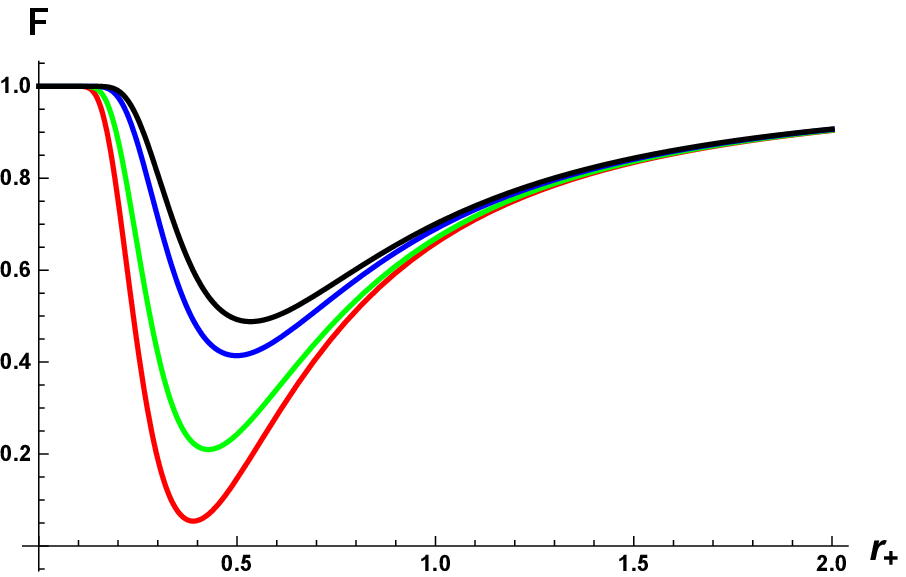}\\
{\textbf{Figure 8}: Helmholtz free $F$ energy versus $r_{+}$}.
\end{center}
In \textbf{Fig. 8}, it can be observed that the $F$ gradually decreases for the different choices of correction parameter $\eta$. It means that the considered geometry changes its behavior towards equilibrium state, so, we cannot extract further work from it. Generally, the minimum condition of Helmholtz free energy means that there exists an equilibrium state of maximum entropy. 
Moreover, the internal energy ($E=F+T\tilde{S}$) for the new type of BH is computed as  \cite{6}
\begin{eqnarray}\label{2.7}
E&=&-\Big[r_ {\eta} +\eta  \left(r_-r_+\right) \left(\log (16 \pi )-\log \left(\frac{\left(r_+-r_-\right)^2}{r_+^2}\right)\right)+\left(r_+r_-\right)\left(\eta  \left(\log (16 \pi)-\log \left(\frac{\left(r_+-r_-\right)^2}{r_+^2}\right)\right)+2 \pi  r_+^2\right)\nonumber\\&+&2 \pi  r_+^3-4 \pi  r_- r_+^2 \log \left(r_+\right)\Big]\Big[8 \pi  r_+^2\Big]^{-1}.
\end{eqnarray}
\begin{center}
\includegraphics[width=7.1cm]{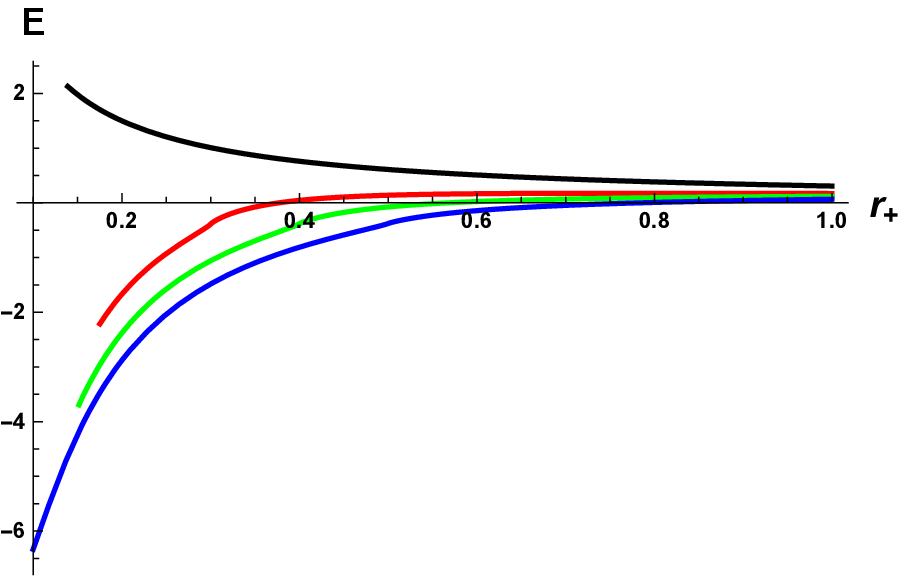}\\
{\textbf{Figure 9}: Internal energy versus horizon radius.}
\end{center}
In \textbf{Fig. 9}, it can be observed that the graph of internal energy depicts negative behavior just before the critical value of radii because of thermal fluctuations. On the other hand, for original case ($\eta=0$), positive behavior of graphs indicate that the BH absorbs heat from the environment. So, its general state depicts that there is a direct proportion relation between internal energy and temperature. Although, the space-time is treated as a thermodynamic system, so, the discussion of pressure and volume ($V=4 \pi r_{+}^{2}$) is much important under these thermal fluctuations. 
So, the BH pressure ($P=-\frac{dF}{dV}$) for the considered geometry takes the form
\begin{equation}\label{2.8}
P=\frac{1}{64 \pi ^2 r_+^4}\left[\left(2 r_--r_+\right) \left(\eta  \left(\log (16 \pi )-\log \left(\frac{\left(r_+-r_-\right)^2}{r_+^2}\right)\right)+2 \pi  r_+^2\right)\right].
\end{equation}
\begin{center}
\includegraphics[width=7.1cm]{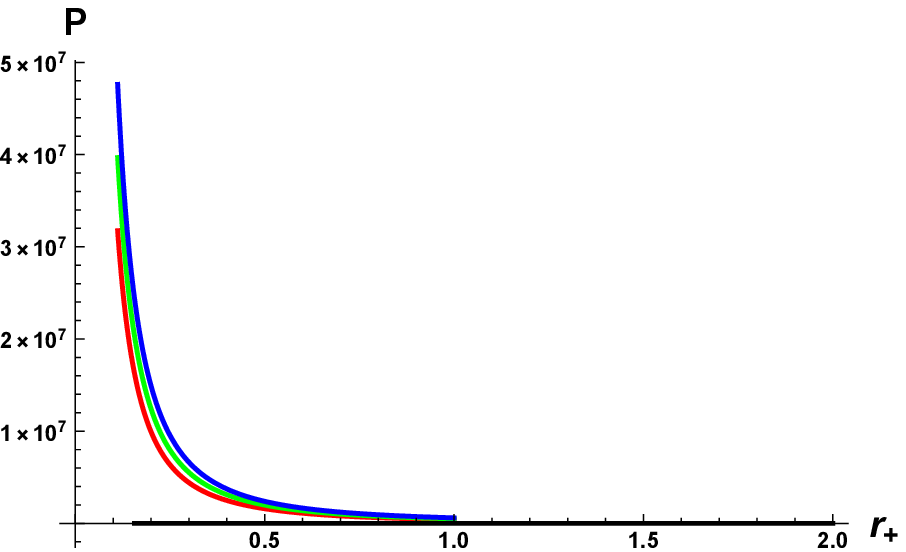}\\
{\textbf{Figure 10}: Pressure versus horizon radius. }
\end{center}
From \textbf{Fig. 10}, it can be observed that $P$ significantly increases for greater values of $\eta$ as well as coincides with the equilibrium state. 

The enthalpy ($H=E+PV$) of the new type of BH can be expressed as
\begin{eqnarray}\label{2.9}
H&=&\Big[-2 \eta  r_-+\eta  r_+ \left(3 \log (16 \pi )-3 \log \left(\frac{\left(r_+-r_-\right)^2}{r_+^2}\right)\right)-2 r_- \nonumber\\&\times&\left(\eta  \left(\log (16 \pi )+2 \pi  r_+^2-\log
   \left(\frac{\left(r_+-r_-\right)^2}{r_+^2}\right)\right)\right)-2 \pi  r_+^3+4 \pi  r_- \nonumber\\&\times&r_+^2 \left(2 \log \left(r_+\right)+1\right)\Big]\Big[16 \pi  r_+^2\Big]^{-1}
\end{eqnarray}
\begin{center}
\includegraphics[width=6.6cm]{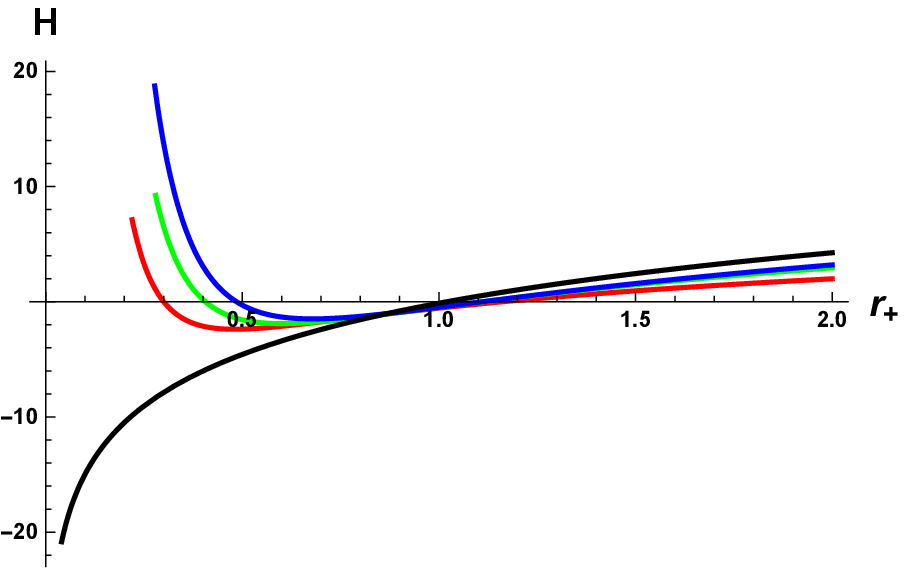}\\
{\textbf{Figure 11}: Enthalpy versus horizon radius. }
\end{center}
In \textbf{Fig. 11}, the enthalpy of the considered system fluctuate along horizon for both corrected and original system. It can be noted that, the enthalpy gradually decreases for large radii of BH, which corresponds to the exothermic reactions. These types of reactions are basically related to heat releasing into its surroundings.

Moreover, the corrected Gibbs free energy ($G=H-T\tilde{S}$) under these fluctuations is calculated as
\begin{eqnarray}\label{2.10}
G&=&\left[-2 \eta  r_-+\eta  r_+ \left(\log (16 \pi )-\log \left(\frac{\left(r_+-r_-\right)^2}{r_+^2}\right)\right)-6 \pi  r_+^3+4 \pi  r_- r_+^2 \left(2 \log
\left(r_+\right)+1\right)\right]\Big[16 \pi  r_+^2\Big]^{-1}.
\end{eqnarray}
\begin{center}
\includegraphics[width=7.1cm]{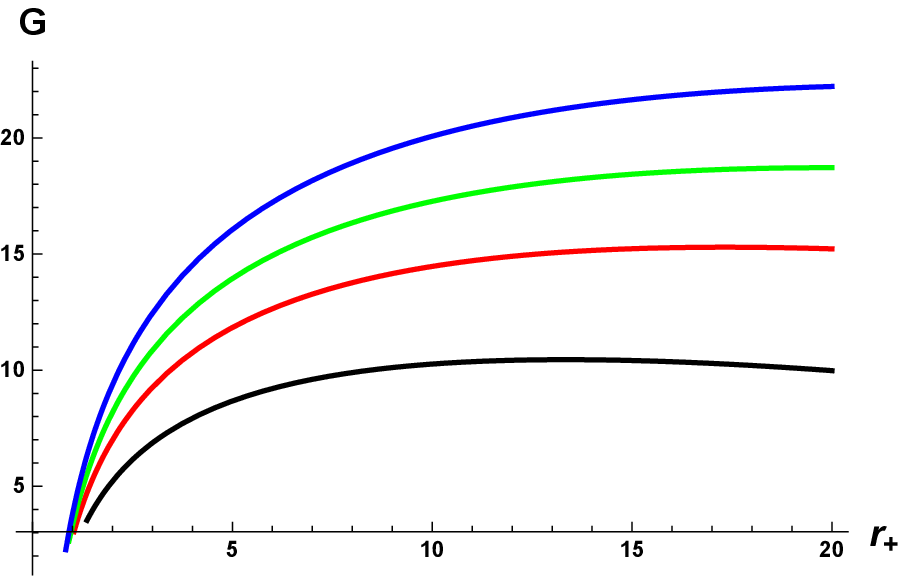}\\
{\textbf{Figure 12}: Gibbs free energy versus horizon radius.}
\end{center}
In \textbf{Fig. 12}, one can observed that for the same previous choices of the correction parameter $\eta$, the $G$ shows positive behavior in the domain $0\leqslant r_+\leqslant20$. This positive behavior of Gibbs free energy indicates that the reaction occurring inside the BHs are non-spontaneous reactions. It means that the system requires more energy to gain equilibrium position. To analyze the stability conditions under these perturbations, the specific heat ($C_{\tilde{S}}=\frac{dE}{dT}$) can be calculated as
\begin{eqnarray}\label{2.11}
C_{\tilde{S}}&=&\Big[\eta  r_+ \left(2 \log \left(\frac{\left(r_+ -r_-\right)^2}{r_+^2}\right)-2 \log (16 \pi )\right)+2 \eta  r_- \left(-\log
\left(\frac{\left(r_+-r_-\right)^2}{r_+^2}\right)-1\right.\nonumber\\&+&\left.\log(16 \pi )\right)+2 \eta r_- \left(\log (16\pi )-\log \left(\frac{\left(r_+-r_-\right)^2}{r_+^2}\right)\right)+4
\pi r_- r_+^2\Big]\Big[4 r_- -2 r_+\Big]^{-1}.
\end{eqnarray}
\begin{center}
\includegraphics[width=7.1cm]{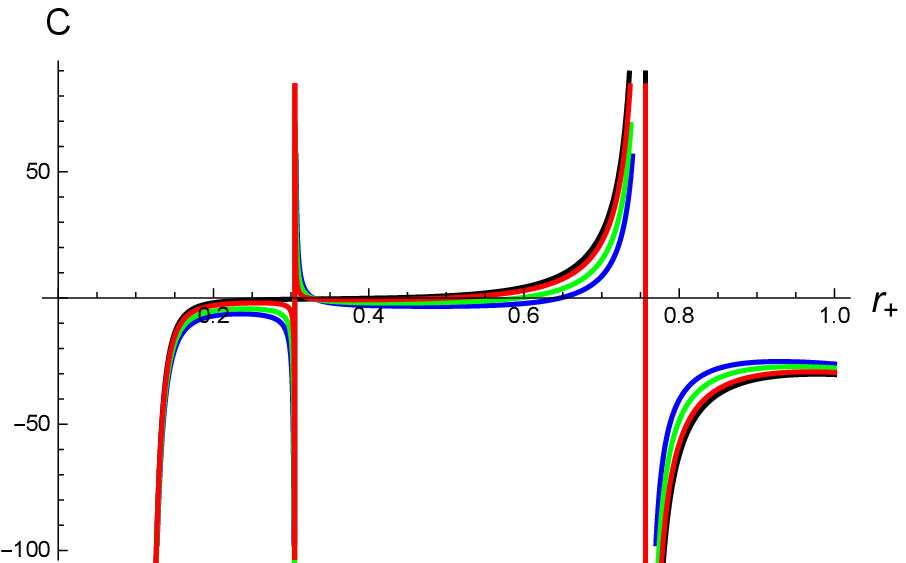}\\
{\textbf{Figure 13}: Specific heat versus horizon radius.}
\end{center}
From \textbf{Fig. 13}, it can be noted that a second order phase transition appears in the given system. The plot of heat capacity $C$ crosses the line of horizon radius at two points like $0.35$ and $0.78$, so, at these points phase transition exists and system lies between stable and unstable phases. Moreover, it can also be observed that the $C$ depicts negative behavior before the occurrence of phase transition, it means that under the effects of thermal fluctuations small BH is unstable while the large one come into stable form after phase transition.

There is another criteria to check the stability of the considered system which is the trace of Hessian matrix. This matrix includes the some thermodynamic quantities like chemical potential ($\upsilon$) derivatives of Helmholtz free energy with respect to hawking temperature. The matrix is defined as
\begin{equation*}\nonumber
\mathcal{\tilde{H}}=\left(
\begin{array}{cc}
\mathcal{\tilde{H}}_{11} & \mathcal{\tilde{H}}_{12} \\
\mathcal{\tilde{H}}_{21} & \mathcal{\tilde{H}}_{22} \\
\end{array}
\right),
\end{equation*}
where
\begin{equation}\nonumber
\mathcal{\tilde{H}}_{11}=\frac{\partial^{2} F}{\partial T^{2}},\quad
\mathcal{\tilde{H}}_{12}=\frac{\partial^{2} F}{\partial T \partial \upsilon},\quad
\mathcal{\tilde{H}}_{21}=\frac{\partial^{2} F}{\partial \upsilon \partial T },\quad
\mathcal{\tilde{H}}_{22}=\frac{\partial^{2} F}{\partial \upsilon^{2}}.
\end{equation}
It is noted that the some eigenvalues of this matrix are zero ($\mathcal{\tilde{H}}_{11}\mathcal{\tilde{H}}_{22}=\mathcal{\tilde{H}}_{12}\mathcal{\tilde{H}}_{21}$), so we use the trace to check the stability of the system
\begin{equation}\nonumber
\tau\equiv Tr(\mathcal{\tilde{H}})=\mathcal{\tilde{H}}_{11}+\mathcal{\tilde{H}}_{22},
\end{equation}
where
\begin{eqnarray}\nonumber
\mathcal{\tilde{H}}_{11}&=&\left[2 r_+^2 \left(\eta  \left(\log (16 \pi )-\log \left(\frac{\left(r_+-r_-\right)^2}{r_+^2}\right)\right)+5 \pi  r_-^2\right)-\eta  r_- r_+ \left(5 \log (16 \pi )-5 \log
   \left(\frac{\left(r_+-r_-\right)^2}{r_+^2}\right)\right)\right.\\\nonumber&+&\eta  r_-^2 \left(-3 \log \left(\frac{\left(r_+-r_-\right)^2}{r_+^2}\right)+3 \log (16 \pi
   )+2\right)+6 \pi  r_+^4-16 \pi  r_-\left. r_+^3\right]\Big[2 \left(r_+-3 r_-\right)\left(r_--r_+\right)\Big]^{-1},
\end{eqnarray}
\begin{eqnarray}\nonumber
\mathcal{\tilde{H}}_{22}&=&\Big[( r_+ r_-)^{-1}(2 r_+- r_-)^{2}\Big]\Big[2\left(r_+-3 r_-\right) \left(r_--r_+\right)\Big]^{-1}\left[2 r_+^2 \left(5 \pi r_-^2+\eta  \left(\log (16 \pi )-\log \right. \left(\frac{\left(r_+-r_-\right)^2}{r_+^2}\right)\right)\right)\\\nonumber&-&\left.\eta  r_- r_+ \left(5 \log (16 \pi )-5 \log
   \left(\frac{\left(r_+-r_-\right)^2}{r_+^2}\right)\right)+\eta  r_-^2 \left(-3 \log \left(\frac{\left(r_+-r_-\right)^2}{r_+^2}\right)+2+3 \log (16 \pi
   )\right)+6 \pi  r_+^4\right.\\\nonumber&-&16 \pi \left. r_- r_+^3\right].
\end{eqnarray}
Generally, the criteria to be a system in stable form is the positive behavior of the trace of the matrix like $Tr(\mathcal{H})\geq0$. 
\begin{center}
\includegraphics[width=7.1cm]{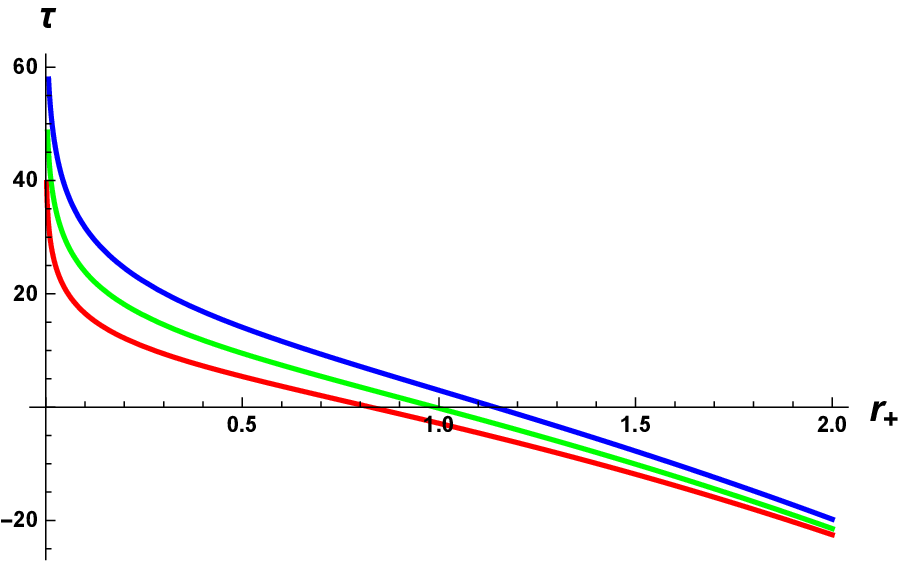}\\
{\textbf{Figure 14}: Gibbs free energy versus horizon radius. }
\end{center}
From (\textbf{Fig. 14}), it can be seen that for the small variation of horizon radius, graph depicts positive behavior, therefore, stability exist. But generally small BH looks unstable as the behavior of graph is gradually decreases. So, here, we use two different techniques to discuss the stability of the system. By comparison, it is noticed that the thermal fluctuations effects the stability of small radii BH.
\section{Conclusions}
In this article, the thermodynamic quantities for new version of Schwarzschild BH has been computed. Firstly, we have studied the Hawking temperature through entropy and then, we have calculated the density of state with the help of inverse Laplace transformation. We have concluded from our graphical analysis that for larger choices of outer horizon, the graph gradually decreases and the energy released from the system at this condition.
Moreover, we have derived the Hawking temperature with the help of surface gravity and analyzed the graphical behavior of temperature. We have observed that the decreasing temperature with increasing horizon completely satisfies the Hawking's phenomenon and temperature at a peak with non-zero horizon states the condition of BH remnant. For the thermodynamic parameter $p=0$ and $\mu=2M$ in Eq. (\ref{bb}), we get the Hawking temperature of Schwarzschild BH given in Eq. (\ref{dd}) for $\hbar=k=G=1$.

 We have observed the exact entropy of the system depends on Hawking temperature by taking into account the steepest descent method under the conditions $\frac{\partial \tilde{S}}{\partial\beta}=0$ and $\frac{\partial^{2} \tilde{S}}{\partial\beta^{2}}>0$. Furthermore, we have observed the monotonically increasing behavior of entropy in the domain $0\leqslant r_+\leqslant1$. We have concluded that, the height of entropy decreases for a specific value of the radii, that means the abrupt change represents a thermal fluctuation in the system.
 We have also studied that the small BH is more effective under thermal fluctuations. 
 
 Moreover, we have studied the corrected entropy to analyze the effects of thermal fluctuations for the corresponding BH and also evaluated the thermodynamic quantities i.e., internal energy, Helmholtz free energy, Gibbs free energy, enthalpy, specific heat and pressure. We have observed that the large BH is stable but the small BHs get unstable regions due to first-order corrections.

In our graphical analysis, it is noted that the Helmholtz free energy gradually decreases for the different choices of correction parameter $\eta$. This means that the considered geometry change its behavior towards equilibrium state, so, we cannot extract further work from it. Generally, the minimum condition of Helmholtz free energy means that there exists a equilibrium state of maximum entropy. 

The internal energy ($E=F+T\tilde{S}$) for the new type of BH has been evaluated. It has been observed that the graph of internal energy depicts negative behavior just before the critical value of radii because of thermal fluctuations. On the other hand, for ($\eta=0$), positive behavior of graphs indicate that the BH absorbs heat from the environment. So, its general state ensures that there is a direct proportion relation between internal energy and temperature. The metric is treated as an important thermodynamic of pressure and volume under these thermal fluctuations. In the original and corrected enthalpy system, we have noted that the enthalpy gradually decreases for large radii of BH and which associates to the exothermic reactions and these type of reactions are basically related to heat releasing into its environment.

In the particular domain $0 \leq r_+\leq 20$, the Gibbs free energy showed positive behavior for different values of correction parameter $\eta$ and it implies that the reaction occurring into the inner part of BH is non-spontaneous reactions and they required more energy to get its equilibrium position. At the end, we checked the stability of the system with the help of two different techniques specific heat and Hessian matrix. By comparison of these methods, we have concluded that the specific heat depicts negative behavior before the occurrence of phase transition, it means that under the effects of thermal fluctuations small BH is unstable while the large one comes into stable form after phase transition.
\section*{Appendix A}
\begin{center}
\includegraphics[width=7.1cm]{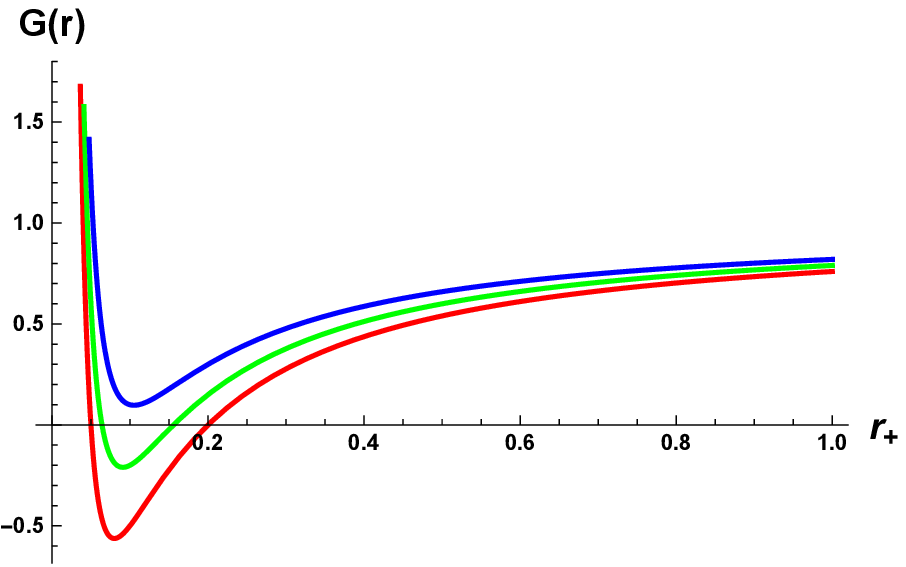}\\
{\textbf{Figure 1}: Metric function versus $r_+$ for $r_-= 0.7$ (red), $r_-=0.8$ (green) and $r_-=0.9$ (blue) . }
\end{center}
\begin{center}
\includegraphics[width=7.1cm]{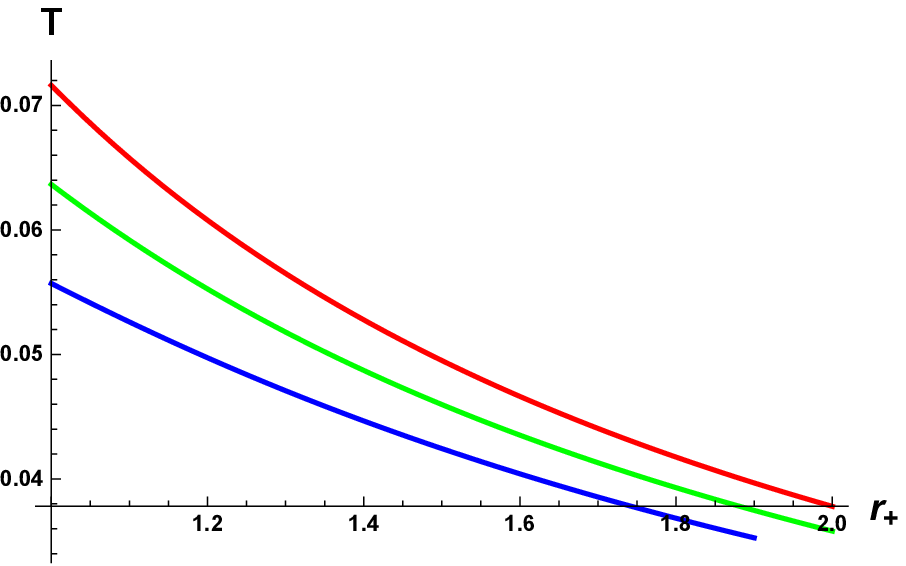}\\
{\textbf{Figure 2}: Hawking temperature versus $r_+$. }
\end{center}
\section*{Appendix B}
\begin{center}
\includegraphics[width=7.1cm]{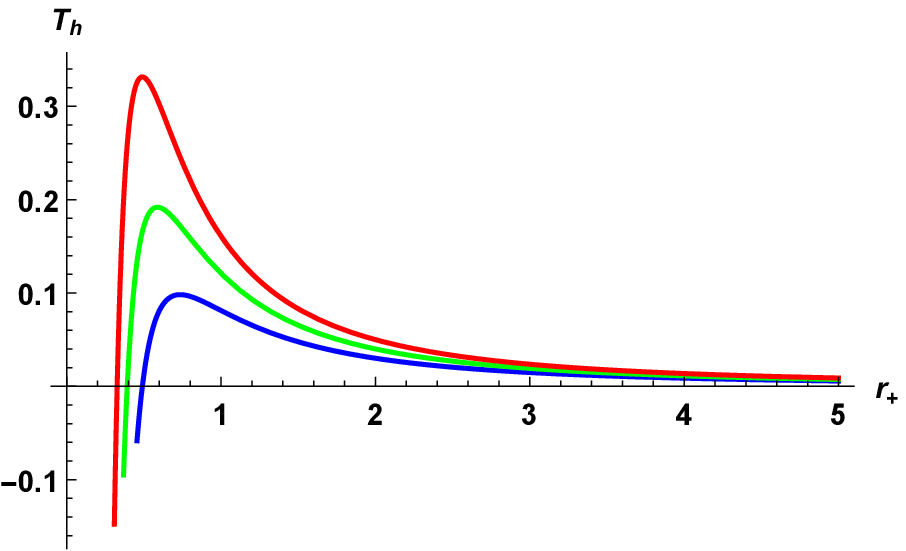}\\
{\textbf{Figure 5}: $T_{h}$ versus $r_+$ for~$\mu= 2$ (blue), $\mu=2.5$
(green), $\mu=3$ (red) and fixed $p=0.7$. }
\end{center}
\begin{center}
\includegraphics[width=7.1cm]{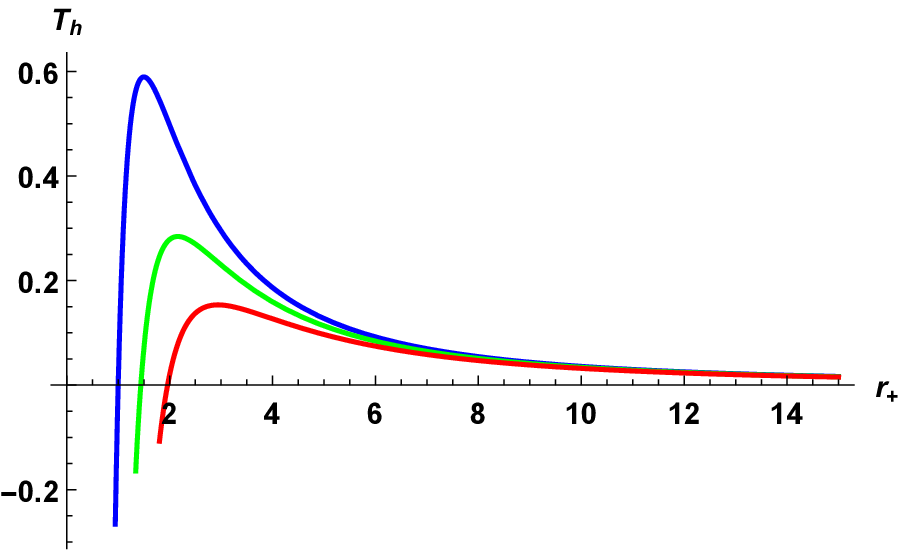}\\
{\textbf{Figure 6}: $T_{h}$ versus $r_+$ for~$p= 5$ (blue), $p=6$
(green), $p=7$ (red) and fixed $\mu=50$. }
\end{center}

{\bf DATA AVAILABILITY}\\
The data underlying this article will be shared on reasonable request to the corresponding authors.

\end{document}